\documentclass[12pt,thmsa]{article}
\usepackage{sw20lart}

\input tcilatex
\QQQ{Language}{
British English
}

\begin{document}

\title{Cosmologies with Varying Light-Speed}
\author{John D. Barrow \\
Astronomy Centre,\\
University of Sussex\\
Brighton BN1 9QJ\\
UK}
\maketitle
\date{}

\begin{abstract}
We analyse a generalisation of general relativity that incorporates a cosmic
time-variation of the velocity of light in vacuum, $c,$ and the Newtonian
gravitation 'constant', $G,$ proposed by Albrecht and Maguejo. We find exact
solutions for Friedmann universes and determine the rate of variation of $c$
required to solve the flatness and classical cosmological constant problems.
Potential problems with this approach to the resolution of the flatness and
classical cosmological constant problems are highlighted. Reformulations are
suggested which give the theory a more desirable limit as a theory of
varying $G$ in the limit of constant $c$ and its relationship to theories
with varying electron charge and constant $c$ are discussed.\\ 

\end{abstract}

\section{Introduction}

Albrecht and Magueijo \cite{alb} have recently drawn attention to the
possible cosmological consequences of time variations in $c,$ the velocity
of light in vacuum. In particular, there are mathematical formulations of
this possibility which offer new ways of resolving the horizon and flatness
problems of cosmology, distinct from their resolutions in the context of the
standard inflationary universe theory \cite{infl} or the pre-Big-Bang
scenario of low-energy string theory \cite{prebb}. Moreover, in contrast to
the case of the inflationary universe, varying $c$ may provide an
explanation for the relative smallness of the cosmological constant today.
The study presented in ref.\cite{alb} envisages a sudden fall in the speed
of light, precipitated by a phase transition or some shift in the values of
the fundamental constants, and explores the general consequences that might
follow from a sufficiently large change.

In this paper we explore the formulation of ref.\cite{alb} in further
detail, showing that one can solve the cosmological field equations that
define it for general power-law variations of $c$ and $G$. This allows us to
determine the rate and sense of the changes required in $c$ if the flatness,
horizon, and cosmological constant problems are to be solved. We find that
these questions have simple answers that generalise the conditions for an
inflationary resolution of the flatness and horizon problems in standard
generalised inflation. We will also explore some of the consequences of the
varying $c$ model for other cosmological problems, like isotropy, and its
implications for the evolution of black hole horizon. We will then look more
critically at the formulation of a varying-$c$ theory of gravitation that is
proposed in ref. \cite{alb}, and suggest some ways in which it might be
improved. In particular, since the proposed formulation also includes the
possibility of varying gravitation 'constant', $G(t),$ we consider the
relation between this part of the theory and other well-defined
scalar-tensor gravity theories which incorporate varying $G.$

There have been several high-precision tests of any possible space and time
variations of the fine structure constant at different times in the past, 
\cite{alpha}, \cite{alpha2}. Although any theory which admits $c$ variation
allows us to compare its predictions with the observational constraints on
time variations of the fine structure constant, these observational limits
arise from relatively low-redshift observations or from primordial
nucleosynthesis and are not necessarily of relevance to the varying-$c$
theory under examination here. The period of $c$-variation is expected to be
confined to the very early universe and its observational consequences are
likely to be manifested most sensitively through the spectrum of any
microwave background fluctuations that are created.

\section{The Albrecht-Magueijo Model}

If we begin with the field equations of general relativity

\begin{equation}
G_{ab}-g_{\mu \nu }\Lambda =\frac{8\pi G}{c^4}T_{ab}  \label{ein}
\end{equation}
then we cannot simply 'write in' variations of constants like $G$ or $c$ as
we could in a non-geometrical theory like Newtonian gravity \cite{barG}. For
example, if we wished to allow $G,$ say, to vary in time then we encounter a
consistency problem if the energy-momentum tensor $T_{ab}$ and its vanishing
covariant divergence retains its usual form and physical meaning as
conservation of energy and momentum. The covariant divergence of the
left-hand side of (\ref{ein}) vanishes and if the conservation of energy and
momentum is expressed by the vanishing divergence of $T_{ab}$ in the usual
way then $G$ must be constant. Thus scalar-tensor gravity theories of the
Jordan-Brans-Dicke (JBD) sort \cite{JBD} must derive the variation of $G$
from that of a dynamical scalar field which contributes it own
energy-momentum tensor to the right-hand side of the field equations. An
alternative way of incorporating a varying $G$ or $c$ with a minimum of
change to the underlying theory of gravity is to allow the conventional
energy-momentum tensor to have a non-vanishing divergence. Thus the usual
energy and momentum will not be conserved. Equivalently, one might interpret
this as simply changing the definition of the conserved quantity. Thus,
since the divergence left-hand side of (\ref{ein}) must vanish we require
that the divergence of $G(\mathbf{x})c(\mathbf{x})^{-4}T_{ab}$ vanishes.

\emph{\ }\label{vsl} When discussing any theory in which there is a varying
of some traditional 'constant' of Nature, it is important to recognise that
invariant operational meaning can only be attached to space-time variations
of \textit{dimensionless} constants\textit{.} Variations in dimensional
constants can always be transformed away by a suitable choice of coordinate
frame (change of units); this is discussed in detail in ref. \cite{BM}. We
will be interested in theories in which a time-variation of the fine
structure constant is represented through a variation in $c$. An example
(not unique) of a representation as a theory with varying electron charge, $%
e,$ but constant $c$ has been given by Bekenstein, \cite{alpha2}, and
another by Barrow and Magueijo, \cite{BM}. In varying speed of light (VSL)
theories a varying $\alpha $ is interpreted as $c\propto \hbar \propto
\alpha ^{-1/2}$ and $e$ is constant, Lorentz invariance is broken, and so by
construction there is a preferred frame for the formulation of the physical
laws. In the minimally coupled theory one then simply replaces $c$ by a
field in this preferred frame. Hence, the action remains as \cite{BM} 
\begin{equation}
S=\int dx^4{\left( \sqrt{-g}{\left( {\frac{\psi (R+2\Lambda )}{16\pi G}}+%
\mathcal{L}_M\right) }+\mathcal{L}_\psi \right) }  \label{act}
\end{equation}
with $\psi (x^\mu )=c^4$. The dynamical variables are the metric $g_{\mu \nu
}$, any matter field variables contained in the matter lagrangian $\mathcal{L%
}_M$, and the scalar field $\psi $ itself. The Riemann tensor (and the Ricci
scalar) is to be computed from $g_{\mu \nu }$ at constant $\psi $ in the
usual way. This can only be true in one frame: additional terms in $\partial
_\mu \psi $ must be present in other frames; see refs. \cite{alb}-\cite{BM}
for more detailed discussion.

Varying the action with respect to the metric and ignoring surface terms
leads to 
\begin{equation}
G_{\mu \nu }-g_{\mu \nu }\Lambda ={\frac{8\pi G}\psi }T_{\mu \nu }.
\label{field}
\end{equation}
Therefore, Einstein's equations do not acquire new terms in the preferred
frame. Minimal coupling at the level of Einstein's equations is at the heart
of the model's ability to solve the cosmological problems. It requires of
any action-principle formulation that the contribution $\mathcal{L}_\psi $
must not contain the metric explicitly, and so does not contribute to the
energy-momentum tensor.\emph{\ }

Albrecht and Magueijo (AM) propose that a time-variable $c$ should not
introduce changes in the curvature terms in Einstein's equations in the
cosmological frame and that Einstein's equations must still hold. Thus, $c$
changes in the local Lorentzian frames associated with the cosmological
expansion and is a special-relativistic effect. The resulting theory is
clearly not covariant and so implementation of this idea requires a specific
choice of time coordinate. Choosing that specific time to be comoving proper
time, and assuming the universe is spatially homogeneous and isotropic, so
that there are no spatial variations in $c$ or $G$, leads to the requirement
that the Friedmann equations still retain their form with $c(t)$ and $G(t)$
varying$.$ Thus the expansion scale factor obeys the equations \emph{\ }

\begin{eqnarray}
\frac{\dot a^2}{a^2} &=&\frac{8\pi G(t)\rho }3-\frac{Kc^2(t)\ }{a^2}
\label{fr} \\
\ddot a &=&-\frac{4\pi G(t)}3(\rho +\frac{3p}{c^2(t)})a  \label{ac}
\end{eqnarray}
where $p$ and $\rho $ are the density and pressure of the matter,
respectively, and $K$ is the metric curvature parameter. By differentiating (%
\ref{fr}) with respect to time and substituting in (\ref{ac}), we find the
generalised conservation equation incorporating possible time variations in $%
c(t)$ and $G(t)$,

\begin{equation}
\dot \rho +3\frac{\dot a}a(\rho +\frac p{c^2})=-\rho \frac{\dot G}G+\frac{%
3Kc\dot c}{4\pi Ga^2}.  \label{cons}
\end{equation}

The cosmological constant is a further feature of interest in these
theories. In general relativity we know that there are two equivalent ways
of introducing a cosmological constant into the field equations and their
cosmological solutions. An explicit constant $\Lambda $-term can be added to
the Einstein-Hilbert gravitational lagrangian. Alternatively, a matter field
can be introduced which describes a perfect fluid with a stress satisfying $%
p=-\rho c^2$. In theories other than general relativity these two
prescriptions need not lead to the same contribution to the field equations
and their solutions. For example, in Brans-Dicke theory or other
scalar-tensor theories one can see explicitly that they contribute different
terms to the cosmological evolution equations (see Barrow and Maeda \cite
{omphi}). The matter lagrangian term $\mathcal{L}_M$ will in general also
contain large quantum contributions, so $\mathcal{L}_M\equiv \mathcal{L}%
_M[class]+\mathcal{L}_M[quantum].$

If we examine the action given in eq. (\ref{act}) then we see that there are
two ways in which a cosmological constant can arise in a VSL theory. In our
discussion we shall model the explicit $\Lambda $ term of geometrical origin
in (\ref{act}), or (\ref{field}). This can still be interpreted as a matter
field obeying the equation of state $p_\Lambda =-\rho _\Lambda c^2,$ as
would be contributed by any slowly changing self-interacting scalar field.
We do not address the quantum gravitational or quantum field theoretic
versions of the problem discussed in refs. \cite{wein,lamb} or solve the
problem associated with the presence of quantum contributions described by
any constant term in $\mathcal{L}_M[quantum]$.

If we wish to incorporate such a cosmological constant term, $\Lambda $,
(which we shall assume to be constant) then we can define a vacuum stress
obeying an equation of state

\begin{equation}
p_\Lambda =-\rho _\Lambda c^2,  \label{vac}
\end{equation}
where

\begin{equation}
\rho _\Lambda =\frac{\Lambda c^2}{8\pi G}.  \label{A}
\end{equation}
Then, replacing $\rho $ by $\rho +\rho _\Lambda $ in (\ref{cons}), we have
the generalisation

\begin{equation}
\dot \rho +3\frac{\dot a}a(\rho +\frac p{c^2})+\dot \rho _\Lambda =-\rho 
\frac{\dot G}G+\frac{3Kc\dot c}{4\pi Ga^2}  \label{cons2}
\end{equation}

We shall assume that the remaining matter obeys an equation of state of the
form

\begin{equation}
p=(\gamma -1)\rho c^2(t)  \label{gam}
\end{equation}
where $\gamma $ is a constant.

\section{The Flatness and Lambda Problems}

If we specialise the standard situation in general relativity in which we
set $\dot G\equiv 0\equiv \dot c$ then the conservation equation, (\ref
{cons2}), gives $\rho \propto a^{-3\gamma }$ and the $8\pi G\rho /3$ term
will dominate the curvature term, $Kc^2a^{-2}$ at large $a$ so long as the
matter stress obeys $\rho +3p/c^2<0,$ $\rho +p/c^2\geq 0,$ that is, if

\begin{equation}
0\leq \gamma <\frac 23  \label{inf}
\end{equation}
This is what we shall mean by the \textit{flatness problem.} Since the scale
factor then evolves as $a(t)\propto t^{2/3\gamma }$ if $\gamma >0$ or $\exp
\{H_0t\},H_0$ constant if $\gamma =0,$ we see that it also grows faster than
the proper size of the particle horizon scale ($\propto t$) so long as $%
\gamma <2/3.\ $Thus, a sufficiently long period of evolution in the early
universe during which the expansion is dominated by a gravitationally
repulsive stress with $\rho +3p/c^2<0$ can solve the flatness and horizon
problems. Such a period of accelerated expansion is called 'inflation'.
However, we notice that if a constant vacuum stress, $\rho _\Lambda ,$
associated with a non-zero cosmological constant is added to the right-hand
side of (\ref{cons2}) then, in order to explain why it does not totally
dominate the $8\pi G\rho /3$ term at large $a(t),$ we would need a period
during which the early universe was dominated by an extreme fluid with $%
\gamma <0,$ that is $\rho +p/c^2<0.$ This is problematic because it leads to
contraction $\dot a<0$ and apparent instabilities of the vacuum and of flat
space-time. This is what we shall mean by the \textit{cosmological constant
(or lambda) problem}. Other quantum cosmological attempts to solve this
cosmological constant problem have so far proved unsuccessful \cite{lamb}.

We see that the inflationary solution of the flatness problem relies on a
change in the evolutionary behaviour of the matter term in the Friedmann
equation which allows it to dominate the curvature term at large $a(t)$ and
transform the zero-curvature solution into the late-time attractor. In
contrast, the varying-$c$ model of AM will provide a solution of the
flatness and lambda problems by introducing a variation that increases the
rate of fall-off of the curvature ($Kc^2a^{-2}$) and cosmological constant ($%
\rho _\Lambda $) terms in the Friedmann equation with respect to the $8\pi
G\rho /3$ term at large $a(t)$. This is possible for all values of $\gamma $
and does not necessarily require the existence of a period during which $%
\rho +3p/c^2$ is negative.

\section{Exact solutions with Varying $G(t)$ and $c(t)$}

We need to obtain solutions of eqns. (\ref{cons}) and (\ref{cons2}) in order
to evaluate the effects of varying $G$ and $c$ on the expansion dynamics.
Let us first consider the implications for the flatness and horizon problems
alone by setting $\Lambda =0.$

\subsection{The Flatness Problem}

In order to solve eq. (\ref{cons}) we assume that the rate of variation of $%
c $ is proportional to the expansion rate of the universe; that is, 
\begin{equation}
c(t)= c_0 a^n; c_0,n \mbox{ constants.}  \label{c}
\end{equation}
We shall see that we do not need to prescribe a form for the time variation
of $G(t).$

{}From eqs. (\ref{cons}) and (\ref{gam}), we have

\[
\frac{(\rho a^{3\gamma })^{\cdot }}{\rho a^{3\gamma }}=-\frac{\dot G}G+\frac{%
3Kc\dot c}{4\pi G\rho a^2}, 
\]
hence

\[
\frac{(G\rho a^{3\gamma })^{\cdot }}{G\rho a^{3\gamma }}=\ \frac{3Kc\dot c}{%
4\pi G\rho a^2}. 
\]
and so, using (\ref{c}), we obtain

\begin{equation}
(G\rho a^{3\gamma })^{\cdot }=\frac{3Kc_0^2na^{2n+3\gamma -3}\dot a}{4\pi }
\label{so}
\end{equation}
Integrating, we have the exact solutions

\begin{equation}
G(t)\rho a^{3\gamma }=\frac{3Kc_0^2na^{2n+3\gamma -2}}{4\pi (2n+3\gamma -2)}%
+B,\mbox{ if }2n+3\gamma \neq 2,  \label{sol1}
\end{equation}

\begin{equation}
G(t)\rho a^{3\gamma }=\frac{3Kc_0^2n\ln a\ }{4\pi }+B,\mbox{ if }2n+3\gamma
=2,  \label{sol2}
\end{equation}
with a constant of integration, $B>0$. This holds for arbitrary $G(t)$
variations. Analogous solutions can be found from (\ref{so}) for other
possible variations of $c$ with respect to $a(t)$ or $t$ but we shall
confine our attention to the simple case of eq. (\ref{c}).

Returning to the Friedmann eqn., (\ref{fr}), we see that the condition for
the solution of the flatness problem is the same as that which would hold if 
$G$ were constant. The only impact of $G(t)$ is to multiply $\rho ,$ so we
have in (\ref{fr}), (writing only the $2n+3\gamma \neq 2$ case explicitly)

\begin{eqnarray}
\frac{\dot a^2}{a^2} &=&\frac{2Kc_0^2na^{2n-2}}{(2n+3\gamma -2)}+B^{\prime
}a^{-3\gamma }\ -Kc_0^2a^{2n-2}  \nonumber \\
\frac{\dot a^2}{a^2} &=&B^{\prime }a^{-3\gamma }+\frac{Kc_0^2a^{2n-2}(2-3%
\gamma )}{(2n+3\gamma -2)};B^{\prime }\mbox{ constant,}  \label{fr1}
\end{eqnarray}
and the $B^{\prime }$ term dominates the curvature ($K)$ terms at large $a,$
so the flatness problem can be solved so long as

\begin{equation}
n\leq \frac 12(2-3\gamma ),  \label{con1}
\end{equation}
where we have noted from (\ref{sol2}) that the flatness problem is also
resolved in the $2n+3\gamma =2$ case. We see that this condition is a
straightforward generalisation of the inflationary condition (\ref{inf})
that obtains when $c$ is constant. We confirm that the variation in $G$ does
not enter this condition in a significant way. Also, the sign of $3\gamma -2$
determines the overall sign of the curvature term in the Friedmann equation.
However, unlike in the constant $c$ case, it is possible to resolve the
flatness problem without appeal to a matter source with $3\gamma -2<0.$ In
particular, we see that in a radiation-dominated universe ($\gamma =4/3$),
the flatness problem is soluble if c$(t)$ falls sufficiently rapidly as the
universe expands, with

\begin{equation}
n\leq -1,  \label{rad}
\end{equation}
while for dust ($\gamma =1$) we require a slower decrease, with

\begin{equation}
n\leq -\frac 12.  \label{du}
\end{equation}

\subsection{The Horizon Problem}

The variation of the expansion scale factor at large $a(t)$ approaches

\begin{equation}
a(t)\propto t^{2/3\gamma }  \label{hor1}
\end{equation}
when the curvature term becomes negligible, as in the situation when $c$ is
constant. The proper distance to the horizon increases as

\begin{equation}
d_h\propto ct\propto a^nt\propto t^{(3\gamma +2n)/3\gamma }  \label{hor2}
\end{equation}
and so we see that the scale factor can grow as fast, or faster, than $d_h$
as $t$ increases if

\[
2\geq 2n+3\gamma 
\]

This is the condition to solve the horizon problem and is identical to that
for the solution of the flatness problem, just as in the situation with
constant $c.$

\subsection{The Lambda Problem}

If we consider the case with a non-zero $\Lambda $ term, that is contributed
by a stress obeying $p_\Lambda =-\rho _\Lambda c^2$, then we need to solve (%
\ref{cons2}) with $c(t)$ varying as (\ref{c}). In this case we need to
assume a form for the variation of $G$ with $a$ or $t$, and we shall again
assume that its rate of variation is proportional to the expansion rate of
the universe, with

\begin{equation}
G(t)=G_0a^q  \label{G}
\end{equation}
where $G_0$ and $q$ are constants. We assume that $\Lambda $ is constant.
{}From eq. (\ref{A}) we have

\[
\dot \rho _\Lambda =\frac{\Lambda c_0^2(2n-q)a^{2n-q-1}\dot a}{8\pi G_0} 
\]

Now we look for solutions of (\ref{cons2}) with

\begin{equation}
\rho a^{3\gamma }=Da^{-f}+Ea^{-g}  \label{rho1}
\end{equation}
where $B,D,$and $E$ are constants.

This requires

\begin{eqnarray}
&&\   \nonumber  \label{c1} \\
f &=&q-2n-3\gamma  \label{c2} \\
g &=&2+q-2n-3\gamma =2+f  \label{c3} \\
D &=&\frac{\Lambda c_0^2(q-2n)}{8\pi G_0(2n+3\gamma )}  \label{c4} \\
E &=&\frac{3Kc_0^2n}{4\pi G_0(2n+3\gamma -2)}  \label{c5}
\end{eqnarray}
where we only write down the solution for the cases where $2n+3\gamma \neq 0$
or $2$. Thus, the density is

\begin{equation}
8\pi G(t)\rho =\frac{\Lambda c_0^2(q-2n)a^{2n}}{\ 2n+3\gamma }+\ \frac{%
6Kc_0^2na^{2n-2}}{\ 2n+3\gamma -2}  \label{main2}
\end{equation}
If we substitute in the Friedmann equation containing $\rho _\Lambda $,

\begin{equation}
\frac{\dot a^2}{a^2}=\frac{8\pi G}3(\rho +\rho _\Lambda )-\frac{Kc^2}{a^2}
\label{main3}
\end{equation}
then we have

\begin{equation}
\frac{\dot a^2}{a^2}=\frac{\Lambda c_0^2a^{2n}(q+3\gamma )}{3(\ 2n+3\gamma )}%
+\frac{Kc_0^2a^{2n-2}(2-3\gamma )}{\ 2n+3\gamma -2}  \label{la}
\end{equation}

We see that the $\Lambda $ term on the right-hand side of eq. (\ref{la})
will fall off faster than the curvature and matter density terms so long as $%
c(t)$ falls off fast enough, with

\begin{equation}
n<-\frac{3\gamma }2  \label{con2}
\end{equation}
As expected, this is requires a more rapid increase in $c(t)$ than is
required to solve the flatness or horizon problems. In particular, in the
radiation and dust cases it requires $c$ to fall as the Universe expands,
with $n<-2$ and $n<-3/2,$ respectively.

\section{Better Formulations}

The most challenging problem besetting any attempt to examine the
cosmological consequences of varying $c$ is the formulation of a
self-consistent theory which incorporates such a variation. The AM approach
is not the only way of proceeding and so we should look more critically at
its relation to other things that we know. The most interesting feature in
this respect is the fact that it permits a description of varying $G$ when $%
c $ is kept constant. However, it is clear that in this case the resulting
cosmological equations (\ref{fr}), (\ref{ac}) and (\ref{cons}) are not the
equations describing the evolution of the Friedmann universe in a
scalar-tensor gravity theory, like JBD. The reasons are clear. JBD theory
meets the covariant divergence constraints imposed by the Bianchi identities
by including the energy density contributed by the scalar field that acts as
the source of the variations in $G$. These contributions are not included in
the AM formulation. In order to improve upon it we could take a
scalar-tensor gravity theory and carry out the same procedure used to derive
(\ref{cons}) and check that the time variation of $c$ is still permitted.
This means that we assume that the JBD equations hold for the Friedmann
model, with $c$ allowed to be a time variable, and then derive the
conservation equation from them. For JBD theory we have \cite{coles}, \cite
{omphi},

\begin{eqnarray}
\frac{\dot a^2}{a^2} &=&\frac{8\pi \rho }{3\phi }-\frac{\dot \phi \dot a}{%
\phi a}+\frac{\omega \dot \phi ^2}{6\phi ^2}-\frac{Kc^2(t)\ }{a^2},
\label{frbd} \\
\frac{3\ddot a}a &=&\ -\frac{8\pi }{(3+2\omega )\phi }\left[ (2+\omega )\rho
+\frac{3(1+\omega )p}{c^2(t)}\right] -\ \frac{\ddot \phi }\phi -\frac{\omega
\dot \phi ^2}{\phi ^2},  \label{acbd}
\end{eqnarray}

\begin{equation}
\ddot \phi +\frac{3\dot a\dot \phi }a=\frac{8\pi }{3+2\omega }(\rho -\frac{3p%
}{c^2(t)})=\frac{8\pi \rho (4-3\gamma )}{3+2\omega }.  \label{phi}
\end{equation}
{}From these three equations we obtain the generalised conservation
equation. Since after we impose the equation of state, the varying $c$ term
only appears in the first of these equations, the only new contribution to $%
2\dot a\ddot a$ is from the $-2Kc\dot c$ term. Hence, we have \cite{joao}

\begin{equation}
\dot \rho +3\frac{\dot a}a(\rho +\frac p{c^2})=\ \frac{3Kc\dot c\phi }{4\pi
a^2}.  \label{cons2a}
\end{equation}
It is instructive to compare this to the AM equation (\ref{cons}). We see
that the incorporation of $G$ variation differs. The $\rho \dot G/G$ term of
(\ref{cons2a}) is absent but the $K$ term has the same form with $G(t)$
replaced by $\phi ^{-1}(t)$ as usual in JBD theory. However, the new system
is more constrained because solutions to eqns. (\ref{phi}) and (\ref{cons2a}%
) must be consistent.

When the matter content is radiation the conservation equation (\ref{phi})
has the particular solution 
\[
\phi =G^{-1}=constant 
\]
and then (\ref{cons2a}) is identical to the AM equation (\ref{cons}) in the
case where $G$ is constant but $c$ varies with time. The conditions for
resolving the flatness and cosmological constant problems then become the
same as given for the AM formulation above. For other equations of state the
situation is more complicated. Rearranging the equations, we need to solve

\[
(\rho a^{3\gamma })^{\cdot }=\frac{3Kc\dot c\phi a^{3\gamma -2}}{4\pi } 
\]

\[
(\dot \phi a^3)^{\cdot }=\frac{8\pi \rho a^3(4-3\gamma )}{3+2\omega } 
\]
together with eq. (\ref{frbd}). When $K=0$ they are solved by the usual
exact solutions of JBD theory. When the universe is radiation dominated
these equations have particular solutions with $\phi $ constant which are
identical to those derived in the last section for the case of varying $c$
and constant $G.$ These equations will be studied in more detail elsewhere 
\cite{BM2}.

The JBD theory is not the most general scalar-tensor gravity theory. It is a
particular case of a general class in which the constant BD parameter, $%
\omega $, is a function of the scalar field $\phi ,$ \cite{omphi}$.$ For
these theories, the Friedmann universe (with $c$ constant) is governed by
the equations (in the $\Lambda =0$ case)

\[
2\left( \frac{\dot a}a\right) ^{\cdot }+\left( \frac{\dot a}a\right) ^2+2%
\frac{\dot a\dot \phi }{a\phi }+\frac{Kc^2}{a^2}=-\frac{\omega (\phi )\dot
\phi ^2}{2\phi ^2}-\frac{8\pi p}{\phi c^2}-\frac{\ddot \phi }\phi , 
\]
$\ $

\begin{equation}
\frac{\dot a^2}{a^2}=\frac{8\pi \rho }{3\phi }-\frac{\dot a\dot \phi }{a\phi 
}+\frac{\omega (\phi )\dot \phi ^2}{6\phi ^2}-\frac{kc^2}{\ a^2},
\label{frst}
\end{equation}

\begin{equation}
\ddot \phi +3H\dot \phi +\frac{\omega ^{\prime }\dot \phi ^2}{2\omega +3}=%
\frac{8\pi \rho (4-3\gamma )}{2\omega +3}.  \label{fist}
\end{equation}

{}From these equations we obtain the same generalisation of the mass-energy
conservation as in JBD (which is the case $\omega (\phi )=$ constant), given
by eq. (\ref{cons2a}). Again, radiation-dominated solutions (with $\phi
=G^{-1}$ constant) and varying $c\ $of the form found in the last section
are always exact particular solutions of these equations.

\section{Some Problems}

\subsection{The Velocity Problem}

If $c$ varies here may be a problem with the perturbations to the isotropic
expansion of the universe which are powers of $v/c$. For example, for
rotational velocity perturbations in matter obeying $p=(\gamma -1)\rho c^2,$
the conservation of angular momentum gives $\rho a^4v=$ constant. Although
the original formulation described above does not allow us to impose energy
conservation in the usual way ($E\propto c^2$), we shall assume that angular
momentum conservation is preserved since it does not incorporate direct $c$
dependence. Hence, we have 
\[
v\propto a^{3\gamma -4}. 
\]
Now, if

\[
c(t)=c_0a^n 
\]
we have

\[
\frac vc\propto a^{3\gamma -4-n} 
\]
If the flatness problem is to be solved, we need $2-2n>3\gamma $ so we will
have $3\gamma -4-n\ <-2-3n.$ For the $\Lambda $ problem to be solved we need 
$3\gamma +2n<0.$ Hence, in the radiation era we need $n<-1$ to solve the
flatness problem and $n<-2$ to solve the $\Lambda $ problem. Therefore, when 
$\gamma =4/3,$we have

\[
\frac vc\propto a^{-n} 
\]
we see that $v/c$ grows in time in both cases as $a\rightarrow \infty $.
During the dust era, $\gamma =1,$ and so

\[
\frac vc\propto a^{-1-n} 
\]
where $n<-1/2$ solves flatness (allowing $v/c$ to decay in $-1<n<-1/2$) and $%
n<-3/2$ solves $\Lambda $ (which would always require $v/c$ to grow)$.$

In general, the condition for solution of the flatness problem has $v/c$
growing as a power of $a$ equal to $3\gamma -4-n>9\gamma /2-5,$ while a
solution of the $\Lambda $ problem has it growing as $3\gamma -4-n>9\gamma
/2-4.$ So we can only \textit{avoid} blow-up of the velocity perturbations
at large $a$ if $\gamma <10/9$ (if we also want to solve flatness) or $%
\gamma <8/9$ (if we also want to solve $\Lambda $).

\subsection{Shear Evolution}

We can extend the AM model to include the effects of simple anisotropies
with isotropic curvatures. This would correspond to generalisations of the
Bianchi type I and type V universes in the flat and open cases,
respectively. In the spirit of equations (\ref{fr})-(\ref{ac}) the required
generalisation which includes the shear anisotropy scalar, $\sigma $, is

\begin{equation}
\frac{\dot a^2}{a^2}=\frac{8\pi G(t)\rho }3+\frac{\sigma ^2}3-\frac{Kc^2(t)\ 
}{a^2}  \label{sh1}
\end{equation}

\begin{equation}
\frac{\ddot a}a=-\frac{4\pi G(t)}3(\rho +\frac{3p}{c^2(t)})a-\frac{2\sigma ^2%
}3  \label{sh2}
\end{equation}
where $a(t)$ is now the geometric-mean expansion scale factor. From these
equations we may derive the analogue of the matter conservation equation:

\begin{equation}
\dot \rho +3\frac{\dot a}a(\rho +\frac p{c^2})+\frac \sigma {4\pi G}(\dot
\sigma +3\frac{\dot a}a\sigma )=-\rho \frac{\dot G}G+\frac{3Kc\dot c}{4\pi
Ga^2}.  \label{sh3}
\end{equation}

This equation resembles the conservation equation for two non-interacting
perfect fluids, one of which (the 'anisotropy' energy density) has the
equation of state $p=\rho c^2$. However, we see that the original AM
prescription leaves the material density coupled to the shear evolution. If
we add the shear evolution equation from the standard (constant $c$) case
then we have

\begin{equation}
\dot \sigma +3\frac{\dot a}a\sigma =0  \label{sh4}
\end{equation}
and the density evolution obeys the same equation as that for the isotropic
model ($\sigma =0$).

\subsection{Black Holes}

It is interesting to note that if a black hole forms with radius

\begin{equation}
R_g=\frac{2GM}{c^2}  \label{den}
\end{equation}
then, from (\ref{den})

\[
\frac{\dot R_g}{R_g}=\frac{\dot G}G-2\frac{\dot c}c=-\frac{\dot \rho
_\Lambda }{\rho _\Lambda }. 
\]
This seems to imply that if $G$ is constant and $c$ is falling then
primordial black holes would grow in size at the same rate as the universe
is expanding. There will also be consequences for black hole evaporation.
The Planck scale will grow rapidly $\propto c^{-5}$. It is possible to
examine the changes to the laws of black hole mechanics using the results
for the change in the characteristics of black body radiation thermodynamics
but we do not digress to discuss this here.

\subsection{Late-time fields}

Another aspect of this model is worth commenting on. We have determined the
rate of variation of $c(t)$ that would allow the flatness (or $\Lambda $) to
be solved. However, we could also use these equations in the cases where the
flatness problem was \textit{not }solved in the same way as a description of
a cosmological matter field that plays the role of a matter field which can
dominate the expansion dynamics of the universe at late times. Of course, as
with the possibility of present-day contributions to the dynamics by the
curvature or $\Lambda $ terms in the case where $c$ is constant, this
requires a special tuning of the initial sizes of these terms in the
Friedmann equation with respect to the density term in order that their
effects just start to become significant close to the present epoch.

An interesting point to notice about $c$-varying models is that if the
universe were to contain a dynamically significant cosmological constant
term at late times then it would naturally have negligible curvature term
and so resemble a zero-curvature universe with subcritical total density,
just like inflationary universes with a residual lambda term. This property
allows natural almost flat asymptotes to arise, as pointed out in ref. \cite
{BM2}.

\subsection{Fine structure constant variations}

One interesting feature of any self-consistent theory of $c$ variation is
that it enables us to evaluate the consequences of time (and space)
variations in dimensionless coupling constants like the fine structure
constant. In an expanding universe with $c$ varying we require $\rho
c^2\propto (k_BT)^4/(\hbar c)^3.$ Assuming particle wavelengths are
unchanged in the absence of expansion and the masses of quantum particles
are conserved so $\hbar /c$ remains constant. As the universe expands the
Planck spectrum remain Planckian but the temperature varies as \cite{alb}

\[
\dot T+T\left( \frac{\dot a}a-2\frac{\dot c}c\right) =0. 
\]
Hence,

\[
T\propto \frac{c^2}a\propto a^{2n-1} 
\]
for the power-law variation, (\ref{c}), assumed above. Strong observational
limits could be placed on any deviation of $n$ from zero using data which
establish bounds on the temperature of the microwave background at low
redshifts.

The required variation of $\hbar /c$ means that (for constant $e^2$) the
fine structure constant varies as

\[
\alpha =\frac{e^2}{\hbar c}\propto \frac 1{c^2} 
\]
If we assume a more modest variation in the evolution of $c(t)\propto (\ln
t)^{-1}$ then it becomes possible to examine some possible late-time
consequences of varying $\alpha $ using this model. A detailed particular
theoretical formulation has been provided by Bekenstein, \cite{alpha2}.
Elsewhere, Barrow and Magueijo \cite{BM} have provided a lagrangian-based
theory with varying $e$ which differs from Bekenstein's (unlike Bekenstein's
theory it does not postulate that the gravitational field equations remain
unchanged). This theory may be transformed by a suitable change of units
into a theory with constant electron charge $e$ and varying $c$ like the one
described above, or it may be presented as a theory with constant $c$ and
varying $e$. This features displays explicitly that invariant meaning can
only be ascribed to variations of dimensionless 'constants' (in this case
the fine structure constant).

A more far-reaching aspect of $c$-variation is the fact that a fall in the
value of $c$ at very early times would lead to a strengthening of
dimensionless gauge couplings unless other changes in the structure of
physics could offset this. Thus, attempts to reconstruct the history of the
very early universe could not count on the applicability of asymptotic
freedom and the ideal gas condition at early times in the usual way.

\section{Discussion}

The contemplation of a variation in the speed of light in the very early
universe presents a host of self-consistency problems and boundary effects
with other parts of the core of modern physical theory. Unlike the case of
varying $G$, it requires deep structural changes to many of the foundations
of physics. In this paper we have considered a minimalist varying-$c$ theory
recently proposed by Albrecht and Magueijo \cite{alb} as a new way of
solving the horizon and flatness problems. However, unlike inflation, it
also offers a way of solving the cosmological constant problem. Whereas
Albrecht and Magueijo modelled changes in $c$ as a sudden fall in the
universal value of $c,$occurring as if at a phase transition, we have
considered the behaviour of variations which vary as a power of the
cosmological scale factor in order to determine the rates of change that are
needed to solve different cosmological problems. Our main results are as
follows:

a. If the velocity of light varies with the cosmological scale factor, $%
a(t), $ as $c=c_0a^n$ and the equation of state of matter is $p=(\gamma
-1)\rho c^2,$ then the curvature term becomes negligible for the expansion
of the Universe as $a\rightarrow \infty $ if $c$ falls fast enough, with $%
n\leq \frac 12(2-3\gamma ),$ regardless of the behaviour of $G(t).\ $The
same condition allows for a resolution of the horizon problem. This
generalises the conditions for the inflationary resolution of the flatness
and horizon problems which apply when $c$ is constant $\ (n=0)$ and $G$ is
constant$.$

b. If the velocity of light varies with the cosmological scale factor, $%
a(t), $ as $c=c_0a^n,$the gravitation 'constant' varies as $G(t)=G_0a^q,$
and the equation of state of matter is $p=(\gamma -1)\rho c^2,$ then the
condition for a vacuum stress with $p_\Lambda =-\rho _\Lambda c^2$,
associated with a geometrical cosmological constant term to become
negligible as $a\rightarrow \infty $ is that $c$ fall at a rate such that $%
n<-3\gamma /2.$

c. We have shown how it possible to improve the formulation of the problem
in order to recover a scalar-tensor theory describing $G$ variation in the
limit of constant $c$. This formulation only allows the solutions of the AM
theory to persist in the case when the equation of state is that of black
body radiation. This problem also suggests that we should attempt a fuller
formulation of the varying-$c$ theory by deriving the variation of $c$ from
some scalar field that contributes an energy density to the evolution of the
Universe.

d. We have highlighted some potential problems with the evolution of
rotational velocities as the universe expands. It is also worth noting that
the varying-$c$ model does not solve the flatness and horizon problems by
means of a period of evolution that is close to that described by a
time-translation invariant de Sitter space-time. As a result there does not
appear to be any distinctive set of fluctuations that should emerge from a
period of cosmological evolution in which $c$ changes.

e. Possible variations in $c$ lead to changes in the fine structure constant
and to other gauge couplings in the very early universe if no other changes
in physics exist. However, if the variation in $c$ is confined to a very
early period of evolution soon after the expansion commenced there need be
no conflict with astronomical constraints on any time or space variation in
the fine structure constant.

f. A varying-$c$ theory can be transformed in to a varying-$e$ theory with
constant $c$ by a suitable transformation of units. Theories of this sort
can be given a lagrangian formulation if required.

We believe that these features of the naive varying-$c$ model proposed by
Albrecht and Magueijo are sufficiently interesting for it to be worthwhile
exploring a fuller, more rigorous formulation of a varying-$c$ gravity
theory based upon an action principle which would enable its consequences
for other aspects of physics and cosmology to investigated more rigorously.

\textbf{Acknowledgements}

This work was supported by PPARC and by a Gordon Godfrey Visiting
Professorship at the University of New South Wales, Sydney. I would like to
thank Andy Albrecht and Jo\~ao Magueijo for detailed discussions and for
showing me their results prior to publication, and E. Weinberg for comments.
I would also like to thank John Webb for discussions and hospitality in
Sydney.


\begin{thebibliography}{99}
\bibitem{joao}  Jo\~ao Magueijo has also arrived at this formulation of JBD
theory.
\end{thebibliography}
\end{document}